\documentclass[12pt,letterpaper,hyper]{JHEP3}
\usepackage{amsmath,epsfig}
\usepackage{amssymb,amsfonts}
\usepackage{graphicx}
\usepackage{multirow,cite}
\usepackage{subfigure}

\newcommand{\bi}{\begin{itemize}}
\newcommand{\ei}{\end{itemize}}

\def\p{\partial}

\def\d{\delta}
\def\g{\gamma}
\def\l{\lambda}
\def\e{\epsilon}
\def\k{\kappa}

\def\om{\omega}
\def\s{\sigma}
\def\R0{R^{(0)}}

\def\A{\mathcal{A}}

\def\R{\mathcal{R}}

\def\Om{\Omega}
\def\r{\rightarrow}
\def\half{{\frac12}}

\newcommand{\bea}{\begin{eqnarray}}
\newcommand{\eea}{\end{eqnarray}}
\newcommand{\be}{\begin{equation}}
\newcommand{\ee}{\end{equation}}

\title{A Fefferman-Graham-like expansion for null warped AdS$_3$}
\preprint{}
\author{Monica Guica

\vskip 0.7 cm

{\it Department of Physics and Astronomy, \\ 
\vskip - 0.4 cm \hskip - 0.47 cm \normalsize{University of Pennsylvania,\\ 
\hskip - 0.45 cm  Philadelphia, PA 19104-6396, USA}}}

\abstract{

\vskip 0.7 cm

We consider null warped $AdS_3$ solutions of three-dimensional gravity coupled to a massive vector field. We isolate a certain set of non-propagating solutions to the equations of motion, which we argue are the ones relevant for understanding the stress tensor sector of the dual field theory. We construct a map from these modes to solutions of three-dimensional Einstein gravity with a negative cosmological constant, and thus 
show that they admit a Fefferman-Graham-like asymptotic expansion. We also compute the renormalized on-shell action for these modes at full non-linear level and propose that the dual stress-energy tensor couples to the boundary metric of the auxiliary $AdS_3$ spacetime. The holographic stress tensor we obtain is symmetric, conserved and its trace yields the same conformal anomaly as in $AdS_3$. } 

\begin{document}

\section{Introduction}

The AdS/CFT correspondence \cite{malda,witten} is one of the most profound and yet very concrete insights that we have into the workings of quantum gravity. However, holography is expected to be a general feature of quantum gravity regardless of the background spacetime \cite{thooft}, and it is thus very interesting to try to understand the precise holographic dictionary for other dual pairs of theories. 

A possibly tractable problem is to understand the field theory dual to gravity in null warped $AdS_3$, a spacetime whose metric is given by

\be
ds^2 = - \frac{\l^2 du^2}{\rho^2} +  \frac{ 2\,du \,dv}{\rho} + \frac{\ell^2 d\rho^2}{4 \rho^2} \label{schrintro}
\ee
This spacetime is interesting for at least two reasons. On the one hand, it constitutes
the simplest example of a Schr\"{o}dinger background - that is - a spacetime which geometrically realizes the Schr\"{o}dinger group of non-relativistic scaling symmetries. It is conjectured that gravity in such backgrounds is holographically dual to  strongly-coupled non-relativistic conformal field theories\footnote{Provided one compactifies the null direction $v$, but we will not do so in this article.} \cite{son,balag}. The hope is that we will be able to better understand features of strongly coupled non-relativistic CFTs that occur in condensed matter and cold atom experiments by developing our holographic intuition for these spacetimes. According to this proposed correspondence, gravity in null warped $AdS_3$  should be holographically dual to an intrinsically non-relativistic (and possibly non-local \cite{nr}) two-dimensional theory, which upon compactifying the $v$ direction reduces to a conformal quantum mechanics. 

A different motivation for studying holography in null warped $AdS_3$ has been its potential relevance for understanding the Kerr/CFT correspondence \cite{kerrcft}. More precisely, it has been recently argued that the finite-temperature version of this spacetime appears in the near-horizon of certain extremal higher-dimensional black holes \cite{kerrdip}, and various asymptotic symmetry group analyses indicate the presence of one - and sometimes even two\footnote{Nevertheless, to our knowledge, the boundary conditions proposed so far in the literature only allow for a left-moving Virasoro algebra \emph{or} a right-moving one, but never both simultaneously.} - centrally-extended Virasoro algebras \cite{dipCFT}. The asymptotic symmetry group analyses thus suggest a structure akin to that of \emph{relativistic} two-dimensional conformal field theories.

Usually, the asymptotic symmetries of a given spacetime can be also constructed from
the Brown-York stress tensor or, more generally, the holographic stress-energy tensor of that particular spacetime and the asymptotic Killing vectors \cite{brownyork}. Consequently, we should be able to re-discover the Virasoro algebras seen in the asymptotic symmetry group analyses by constructing the holographic stress energy tensor of the theory.

Before attacking this question, it is useful to remind ourselves of how the $AdS_3$ case works \cite{holoweyl, holorec, vijay}. The Fefferman-Graham theorem ensures that if a metric (asymptotically) satisfies
Einstein's equation

\be
R_{\mu\nu} + \frac{ 2}{\ell^2} \, g_{\mu\nu} =0
\ee
then in radial gauge

\be
ds^2 = \g_{ij}(\rho, x^i) dx^i dx^j + \frac{\ell^2 d\rho^2}{4 \rho^2} \label{radg}
\ee
the induced metric on the $\rho = const.$ slices has an expansion of the form

\be
\frac{\g_{ij}}{\ell^2} = \frac{g^{(0)}_{ij}}{\rho} + g^{(2)}_{ij} + \ldots \label{amads}
\ee
Here $g^{(0)}_{ij}$ is an arbitrary non-degenerate two-dimensional metric. It
determines the trace and covariant divergence of $g^{(2)}_{ij}$ via the asymptotic equations of motion, which read

\be
g^{(2)\, k}{}_k = -\half  R[g^{(0)}] \;, \;\;\;\;\; \nabla_i g^{(2) ij} = \nabla^j  g^{(2)\, k}{}_k \label{aeom}
\ee
The indices are raised and covariant derivatives are taken with respect to the boundary metric $g^{(0)}_{ij}$. The subsequent coefficients in the expansion are recursively determined in terms of $g^{(0)}_{ij}$ and $g^{(2)}_{ij}$. The variation of the renormalized on-shell action with respect to $g^{(0)}$ yields the  expectation value of the holographic stress tensor

\be
T_{ij} = \frac{2}{\sqrt{g^{(0)}}} \frac{\d S[g^{(0)}]}{\d g^{ij}_{(0)}}= \frac{\ell}{8 \pi G} \left(g_{ij}^{(2)} - g^{(0)}_{ij} g^{(2)\,k}{}_k \right) \label{holostress}
\ee
which, due to the asymptotic equations of motion \eqref{aeom}, obeys the expected holographic Ward identities

\be
\nabla_i T^{ij} =0 \;,\;\;\;\;\;\; T^i_i =  \frac{\ell}{16 \pi G} R[g^{(0)}] =   \frac{c}{24 \pi} R[g^{(0)}]  \label{holoward}
\ee
The asymptotic symmetries are those large diffeomorphisms that preserve a flat boundary metric. They consist of boundary diffeomorphisms that can be cancelled by a Weyl rescaling, yielding precisely the two-dimensional conformal group. The conformal anomaly and transformation properties under conformal transformations are precisely those of the stress tensor of a two-dimensional $CFT$ with central charge

\be
c = \frac{3 \ell}{2 G} \label{cc}
\ee
When trying to use a similar procedure to construct the holographic stress tensor in null warped $AdS_3$, one encounters several complications. First, the analogue  of the Fefferman-Graham expansion for null warped $AdS_3$ is not known, and thus one does not have the general form of the asymptotic solution, including the relations between the various coefficients. Second, it is not clear what the scaling properties and conservation laws satisfied by the holographic stress tensor should be. More precisely, if one takes the non-relativistic nature of the system at face value, then one concludes that the dual stress tensor, if chosen to be symmetric, is not conserved. Moreover, certain of its components are irrelevant, certain marginal, and certain relevant with respect to non-relativistic scaling. On the other hand, the asymptotic symmetry group analyses mentioned at the beginning of the introduction point towards a stress tensor obeying conservation laws similar to those obeyed by Lorentz-invariant systems. 

Thus, the set of questions that we would like to address are:

\bi
\item is there a Fefferman-Graham like expansion for null warped AdS$_3$?
\item to which components of the asymptotic metric does the dual stress tensor couple ?
\item what type of conservation laws does the holographic stress tensor obey?
\ei
The question of what the analogue of the Fefferman-Graham expansion is in asymptotically Schr\"{o}dinger spacetimes has been previously addressed in \cite{hartong,ross}. In particular, \cite{hartong} carefully worked out the Fefferman-Graham-like expansion for Schr\"{o}dinger spacetimes obtained from AdS via a TsT transformation, i.e. spacetimes which posess an asymptotically null Killing vector. The main difference with the present work is that in our definition the asymptotically Schr\"{o}dinger spacetime is not required to have any Killing vector, and thus the Fefferman-Graham expansion developed herein is closer in spirit to the one in AdS.

This paper is organised as follows. In section \ref{linm} we describe the setup and discuss the properties of linearized modes in null warped $AdS_3$. In particular, we single out a set of modes, called ``metric modes'', which should be the ones relevant for understanding the stress tensor sector of the dual theory. In section \ref{mapads3}, we construct the most general asymptotic solution for these modes by mapping the problem to $AdS_3$. In section \ref{renact}, we compute the renormalized on-shell action for them and show that it equals the $AdS_3$ answer. We conclude with a discussion and future directions in section \ref{disc}.

\section{Linearized modes in null warped AdS$_3$ \label{linm}}

Null warped $AdS_3$ backgrounds arise as solutions to a variety of theories. We will study the case when they are solutions of the Einstein-Proca theory with a negative cosmological constant. Nevertheless, the results of this paper apply with little modification to any three-dimensional gravitational theory coupled to a massive vector. We start from the action

\be
S = \frac{1}{16 \pi G} \int d^3 x \, \sqrt{g} \, \left(R + \frac{2}{\ell^2} - \frac{1}{4} F_{\mu\nu} F^{\mu\nu} - \half \, m^2 A_{\mu} A^{\mu} \right) \label{massvact}
\ee
The equations of motion are 

\be
R_{\mu\nu} + \frac{2}{\ell^2} \, g_{\mu\nu} - \half F_{\mu\rho} F_{\nu}{}^{\rho} + \frac{1}{4} \, g_{\mu \nu} F^2 - \frac{m^2}{2} \, A_{\mu} A_{\nu} =0 
\ee

\be
\nabla_{\mu} F^{\mu\nu} = m^2 A^{\nu} \label{eqF}
\ee
Taking the divergence of the Maxwell equation yields the constraint $\nabla_{\mu} A^{\mu} =0$, which ensures that the massive gauge field only carries $d =2$ degrees of freedom. If we let

\be
m= \frac{2}{\ell} 
\ee
then the field equations have a solution of the form

\be
d\bar s^2 = - \frac{\l^2}{\rho^2} \, du^2 + \frac{2 du dv}{\rho} + \frac{\ell^2  d\rho^2 }{4 \rho^2} \;, \;\;\;\;\;\;\;\;\bar A = \frac{\l \, du}{\rho}
\ee
For $\l=0$, this spacetime is $AdS_3$ in Poincar\'e coordinates. For $\l \neq 0$, this spacetime is known as null warped $AdS_3$ or three-dimensional Schr\"{o}dinger spacetime, and its isometry group is $SL(2,\mathbb{R}) \times \mathbb{R}_{null}$. For convenience, we will sometimes use the notation $\l_i = \l \, \d^u_i$.

The linearized perturbations around this spacetime have been described in \cite{nr}. After requiring the metric perturbation to be in radial gauge,

\be
g_{\mu\nu} = \bar{g}_{\mu\nu} + h_{\mu\nu} \;, \;\;\;\;\;\;h_{i \rho}= h_{\rho \rho} =0 \;, \;\;\;\;\;\; i,j \in \{u,v \}
\ee
one finds two qualitatively different types of modes, which we denote as $T$-modes and $X$-modes.

\be
h_{ij} = h_{ij}^T + h_{ij}^X \;, \;\;\;\;\; \d A_\mu = \A_\mu^T + \A_\mu^X
\ee
The difference between the $T$-modes  and the $X$-modes is that the first can be induced by radial gauge-preserving three-dimensional diffeomorphisms, whereas the latter cannot. Thus, the T-modes are universal and do not depend on the details of the theory we are studying; the only requirement is that it admit a Schr\"{o}dinger solution.

 The $X$-modes have the characteristic behaviour of scalar fields in Schr\"{o}dinger spacetimes, as the perturbation falls off or blows up with momentum-dependent $(\k_v)$ powers of $\rho$. Meanwhile, the linearized $T$-modes have a relatively simple expansion, involving only integer powers of $\rho$ and logarithms

\be
h_{ij}^T = \frac{h_{ij}^{(-2)}}{\rho^2} + \frac{L_{ij} \ln \rho}{\rho} + \frac{h_{ij}^{(0)}}{\rho} + h_{ij}^{(2)}\;, \;\;\;\;\;
\A_{\mu}^T = \frac{A_{\mu}^{(0)}}{\rho} + A_{\mu}^{(2)} \label{soleom}
\ee
where all the functions that appear depend only on the boundary coordinates $\{ u,v \}$.

The $X$-modes have been discussed extensively in \cite{nr}, where it was also shown how to perform holographic renormalization in the presence of these terms. The main focus of this article are the $T$-modes (which are the ones expected to couple to the dual stress tensor, see arguments in \cite{nr} and at the beginning of section \ref{mapads3}), and we will henceforth set the $X$-modes to zero.

The solution for the metric modes is parametrized by only three functions of both $u$ and $v$, which is the same number as in asymptotically $AdS_3$ spacetimes. Even though we have written the $T$-mode solution in the form of a Fefferman-Graham-like expansion, the coefficients that accompany the various powers of $\rho$ are not determined recursively by the asymptotic equations of motion, as they are in AdS. Moreover, the form of the expansion does receive corrections at higher orders in perturbation theory, for example higher powers of the logarithm appear. One may nevertheless try to study the relations between the various coefficients appearing in \eqref{soleom} using the linearized asymptotic equations of motion. Interestingly, certain relations highly resemble the linearization of  their AdS counterparts

\be
h_{uv}^{(2)} = - \frac{1}{4} \R[h^{(0)}] \;, \;\;\;\;\;\; \p^i h^{(2)}_{ij} = \p^j h^{(2)i}{}_i 
\ee
where $\R[h^{(0)}]$ is the linearized Ricci scalar constructed from $h^{(0)}_{ij}$ 

\be
\R[h^{(0)}] = \p_i \p_j h^{(0)ij} - \p_i \p^i h^{(0)^j}{}_j \label{ricci}
\ee
and we have set $\ell=1$. The relationship between $h^{(-2)}_{ij}$ and the other coefficients is rather complicated, namely

\be
h_{vv}^{(-2)} =0 \;, \;\;\;\;\; h_{uv}^{(-2)} = - \frac{\l^2}{2} h_{vv}^{(0)}\;, \;\;\;\;\;  \R[h^{(-2)}] =-4 \l^2 h_{vv}^{(2)} 
\ee
where $h_{vv}^{(2)}$ is itself non-locally related to $h^{(0)}_{ij}$ and $\R[h^{(-2)}]$ has the same relationship to $h^{(-2)}_{ij}$ as $\R[h^{(0)}]$ to $h^{(0)}_{ij}$ in \eqref{ricci} . This implies a highly non-local relation between $h_{uu}^{(-2)}$ and $h_{uu}^{(0)}$. Finally, the vector field components are simply and locally related to $h^{(-2)}_{ij}$

\be
\l_i A_j^{(0)}+\l_j  A_i^{(0)} = - h^{(-2)}_{ij} \;, \;\;\;\;\; A_{\rho}^{(0)} = - \frac{F}{4}\;, \;\;\;\;\; A_i^{(2)} = - \frac{\p_i F}{4} \;, \;\;\;\; A^{(2)}_{\rho} =0 \label{asymax}
\ee 
where we have defined

\be F \equiv - \half \e^{ij} F_{ij}^{(0)} = \p_u A_v^{(0)} -\p_v A_u^{(0)}
\ee
The remaining asymptotic relations can be found in \cite{nr}.

Given that the relationship between the asymptotic metric components is rather non-local, it is hard to decide what should play the role of source/vev from a holographic point of view. At the same time, a mistake in the identification of the sources is likely to lead to a rather non-sensical theory. Fortunately, there exists an object which contains information about the source/vev organisation, and that is the symplectic form. The symplectic form characterizes the volume element of the phase space of a given system, and is given by

\be
\Om = dp_I \wedge dq^I
\ee
where $I$ runs over the degrees of freedom. In our case, the canonical variables are the boundary metric $\g_{ij}$ and the boundary components of the massive vector field $A_i$.
Here we are working in the radial Hamiltonian formalism, in which the radial distance
from the boundary plays the role of time.  The conjugate momenta are defined via

\be
\d S_{on-shell}  =   \d (S_{bulk} + S_{GH}) \equiv  \int d^2 x (\pi^{ij} \d \g_{ij} +  \pi^i_A \d A_i ) 
\ee
where $S_{GH}$ is the Gibbons-Hawking boundary term. The explicit form of the canonical momenta is

\be
\pi^{ij} =- \frac{\sqrt{\g}}{16 \pi G}   (\, K^{ij}- K \g^{ij}) \;, \;\;\;\;\; \pi^i_A = - \frac{\sqrt{\g}}{16 \pi G}   \,n^{\rho}   F_{\rho}{}^i \d A_i \;, \;\;\;\;\;\;\;\; n^\rho = - 2 \rho
\ee
Consequently, the symplectic form is given by

\be
\Om = \int d^2 x \, ( \d \pi^{ij} \wedge \d \g_{ij} +  \d \pi^i_A \wedge \d A_i ) 
\ee
The symplectic form is finite - given that it is $\rho$-independent - and tells us how the asymptotic solution is organised. Evaluating $\Om$ on the linearized solution \eqref{soleom} and making use of only an \emph{algebraic subset} of the asymptotic linearized equations of motion, one can write it in the following suggestive form\footnote{If we include the $X$-modes, then the symplectic form receives extra contributions of the form $\int X_{source} \wedge X_{vev}$, but there are no cross terms between the $X$-modes and the $T$-modes. We take this as additional circumstantial evidence that the two types of modes can be treated separately.}

\be
\Om=  \frac{\ell}{16 \pi G} \int \left(h_{vv}^{(2)} \wedge h_{uu}^{(0)} - 2 h_{uv}^{(2)} \wedge h_{uv}^{(0)} + h_{uu}^{(2)} \wedge h_{vv}^{(0)}\right)
\ee
Remarkably, this coincides with the symplectic form of linearized perturbations around AdS$_3$ (before the holographic Ward identities are imposed).

One may also  perform holographic renormalization for the linearized metric modes\footnote{In general, one would need to perform holographic renormalization in both sectors simultaneously, since there do exist divergences and counterterms involving both types of terms. Nevertheless, here we will assume without proof that it is consistent to set the $X$-modes to zero. Arguments in favour of this assumption have been given in \cite{marika} in the context of the related chiral scale-invariant models.  }. This consists of computing the on-shell action to second order in the perturbation, regulating it by introducing a small distance cutoff at $\rho=\e$ and subtracting the divergences by a series of covariant counterterms. The needed counterterms  turn out to be remarkably simple

\be
S_{ct} = -\frac{1}{16 \pi G \ell} \int \sqrt{\g} \, \left(2 - A_i A^i - \frac{1}{4} F_{ij} F^{ij} \right) \label{ctact}
\ee
and the variation on the renormalized on-shell action can be written as

\be
\d S_{ren} = \d (S_{on-shell} + S_{ct}) = \half \int  T^{ij} \d h_{ij}^{(0)} 
\ee
where the expression for $T_{ij}$ coincides with the linearization of the holographic stress tensor in $AdS_3$
\be
T_{ij} =\frac{\ell}{8 \pi G} \left( h^{(2)}_{ij} - \eta_{ij} \, h^{(2)k}{}_k \right)
\ee
Such a result is rather intriguing. In the following, we would like to better understand the origin of these AdS-like answers for the symplectic form and the variation of the renormalized action, and whether this behaviour survives at non-linear level. 

\section{Asymptotic expansion of the metric modes \label{mapads3}}

We would like to start this section by characterizing the metric modes in a slightly different manner. In \cite{ross}, it has been observed  that for the background Schr\"{o}dinger solution (in any dimension), a tangent space structure can be chosen such that the massive vector field is proportional to one of the vielbeine. More precisely, if we choose the tangent space metric to be

\be
\eta_{++} = - \l^2 \;, \;\;\;\; \eta_{+-} = \eta_{33} =1
\ee
and the vielbeine as

\be
e^+ = \frac{du}{r} \;, \;\;\;\;\; e^- = dv \;, \;\;\;\;\; e^3 = \frac{dr}{2 r} \;, \;\;\;\;\; 
\ee
then the massive vector field can be written as

\be
A_\mu = \l \, e^+_\mu \equiv \l_a e^a_\mu \label{aviel}
\ee
It has been further argued in \cite{ross}, based on prior work of \cite{ishibashi}, that the most natural way of defining the stress tensor of asymptotically Schr\"{o}dinger spacetimes is by varying the boundary vielbein while keeping the asymptotic tangent space component of the massive vector fixed. As shown in  \cite{ishibashi}, the advantage of such a definition is that the boundary stress tensor resulting from such a variation of the on-shell action with respect to the boundary vielbein satisfies a certain conservation equation. Consequently, if we want to isolate the ``metric modes'', we would like to consider variations of the fields such that the boundary value of the vector field with tangent space indices is fixed, $A_a = \l_a$. 

We immediately note that in three dimensions the metric modes, which correspond to (possibly large) diffeomorphisms of the background solution, do preserve the condition 
$A_\mu = \l_a e^a_\mu$ throughout the  bulk, provided we do not change the tangent space structure. In addition, we note that the background gauge field satisfies a simpler equation than \eqref{eqF}, namely

\be
F_{\mu\nu} = -2 \e_{\mu\nu\l} A^\l \label{eqna}
\ee 
Given that in $3d$ all metric modes are pure gauge, it follows that the generic $T$-modes will also satisfy this simpler equation. Finally, the assumption \eqref{aviel} implies that

\be
A^2 = \l^2 e^+_\mu e^{+\mu} = \l^2 \eta^{++} =0 \label{asq}
\ee
Using the two above simplifications,  Einstein's equations become

\be
R_{\mu\nu} + 2 g_{\mu\nu} = \half (F_{\mu\l} F_\nu{}^\l - \half g_{\mu\nu} F^2) +2 A_\mu A_\nu = 4 A_\mu A_\nu \label{simplein}
\ee
Our goal is now to find the most general solution to the set of equations \eqref{eqna} -\eqref{simplein}.

\bigskip

The fact that $A_\mu \propto e^+_\mu$ imposes a certain constraint on the spin connection associated to the given choice of vielbein. The covariant derivative of the vielbein satisfies

\be
\nabla_{[\mu} e_{\nu]}^+ = - \om_{[\mu}{}^+{}_a\, e_{\nu]}^a = -2 \e_{\mu\nu \l} e^{+\l} 
\ee
A simple way to comply with the above constraint is to require that 
\be
\om^{+a} = -\e^{+ab} e_b \;, \;\;\;\;\; \e_{+-3} =1 \label{condom}
\ee
In general, this is a non-trivial condition, but since it is satisfied in the background spacetime, where

\be
\om^{+-} = \frac{dr}{2r} =  e^3 \;, \;\;\;\;\; \om^{+3} =- \frac{du}{r} =- e^+ 
\ee
it must also be satisfied on any background diffeomorphic to the original one. Thus, \eqref{condom} is a sufficient condition for the existence of a massive vector constructed from the vielbein which obeys the equation of motion \eqref{eqna}. 

We are now interested in the equation of motion obeyed by the metric

\be
\hat{g}_{\mu\nu} = g_{\mu\nu} + A_\mu A_\nu = \hat{\eta}_{ab} \, e_{\mu}{}^a e_\nu{}^b
\ee
where\footnote{Note that $\hat \eta_{ab}$ is just the usual lightcone tangent space metric. }

\be
\hat\eta_{ab} = \eta_{ab} + \l_a \l_b
\ee
For this, we need to understand how to express the Ricci tensor constructed from $\hat{g}_{\mu\nu}$ in terms of the one constructed from $g_{\mu\nu}$. Given that the vielbeine $e_{\mu}{}^a$ are the same for the two metrics, we can compute the difference in the associated spin connections, using the following formula

\be
\om_\mu{}^{ab} = -\half e_{c\mu} (\Omega^{abc} - \Omega^{bca} - \Omega^{cab}) \;,\;\;\;\;\; \Omega_{abc} = e_{a}^\mu e_b^\nu (\p_\mu e_{\nu c} - \p_\nu e_{\mu c})
\ee
We find that

\be
\hat{\Om}^{abc} = \hat{e}^{\mu a} \hat{e}^{\nu b} (\p_\mu e_\nu{}^c - \p_\nu e_\mu{}^c) = \Om^{abc} - \l^a \l_d \,\Om^{dbc} - \l^b \l_d\, \Om^{adc}
\ee
and thus
\be
\hat{\om}_\mu{}^{ab} = \om_\mu{}^{ab} + A_\mu \e^{abc} \l_c
\ee
where we have extensively used \eqref{condom}, as well as the fact that $F_{\mu\nu} A^\nu =0$. After a few manipulations, the Ricci tensor reads

\be
\hat{R}_{\mu\l} = R_{\mu\l} + R_{\mu\nu} A^\nu A_\l - F_{\mu\nu} e^{\nu \,3} A_\l - 2A_\mu A_\l
\ee
Using the fact that

\be
F_{\mu\nu} e^{\nu \, 3} = -2 \l \e_{\mu\nu\rho} e^{\rho +} e^{\nu \,3} = -2 \l \e_{a 3-} e^a_\mu = 2 A_\mu
\ee
we find the following simple relation between the two Ricci tensors

\be
\hat{R}_{\mu\nu} = R_{\mu\nu} + R_{\mu\l} A^\l A_\nu - 4 A_\mu A_\nu
\ee
Plugging this expression into the equations of motion \eqref{simplein}, we find that the equation obeyed by $\hat{g}_{\mu\nu}$ is 

\be
\hat{R}_{\mu\nu} + 2 \hat{g}_{\mu\nu} = 0
\ee
and thus $\hat{g}_{\mu\nu}$ is an AdS$_3$ metric. Consequently, $\hat{g}_{\mu\nu}$ has an asymptotic expansion of the form

\be
ds^2 = \hat \g_{ij} \,dx^i dx^j + \frac{d\rho^2}{4 \rho^2} \;, \;\;\;\;\;\;\; \hat \g_{ij} = \frac{\hat g^{(0)}_{ij}}{\rho} +\hat g^{(2)}_{ij}  + \ldots \label{fg}
\ee 
where

\be
\hat g^{(2) \, i}{}_i = - \half R[\hat g^{(0)}] \;, \;\;\;\;\; \hat\nabla^i \hat g^{(2)}_{ij} = \hat\nabla_j \hat g^{(2)i}{}_i
\ee
Thus, in order to find the most general asymptotic solution for the metric modes in asymptotically null warped $AdS_3$, all we need is to invert the above construction. 
It is not hard to show that the converse computation also works, in the sense that if we start from an asymptotically locally $AdS_3$ solution with a general metric and find a frame such that the spin connection satisfies \eqref{condom} then, letting $A_\mu = \l e_\mu^+$, the metric 

\be
g_{\mu\nu} = \hat{g}_{\mu\nu} - A_\mu A_\nu
\ee
satisfies the Einstein equations of a massive vector theory \eqref{simplein}. The challenge that we face is to find the solution of the equation 

\be
F_{\mu\nu} = - 2 \hat \e_{\mu\nu\l} \hat A^\l \;, \;\;\;\;\; \hat A^2 =0
\ee
in a background $AdS_3$ metric of the generic asymptotic form \eqref{fg}. Assuming a power series expansion for $A_\mu$, we find that

\be
A_\mu = \frac{A_\mu^{(0)}}{\rho} + A_{\mu}^{(2)} +\ldots \label{asya}
\ee
where

\be
A_i^{(0)} = \hat\e_{ij}^{(0)} \hat A^{(0)\,j} \;, \;\;\;\;\; A_\rho^{(0)} = \frac{1}{8} \hat \e^{ij}_{(0)} F^{(0)}_{ij}
\ee

\be
A^{(2)}_i = \hat \e_i{}^{j(0)} \p_j  A_\rho^{(0)} + \hat g^{(2)}_{ij} \hat A^{(0) j} - \frac{1}{2} A_i^{(0)} 
\hat g^{(2) \, k}{}_k \;, \;\;\;\;\;
A_\rho^{(2)} = -\frac{1}{4} \hat\e^{ij}_{(0)} F_{ij}^{(2)} - \frac{1}{2} A_\rho^{(0)} 
\hat g^{(2) \, k}{}_k  \label{arho2}
\ee
Here $\e_{ij}^{(0)}$ is the Levi-Civita tensor density associated with the boundary metric $\hat g^{(0)}$, $F^{(n)} = d A^{(n)}$ and we use a hat for all contravariant tensors whose indices have been raised with $\hat g^{(0)ij}$. The subsequent coefficients in the expansion are also  determined recursively. Given that we are solving a second order differential equation in $\rho$, the general solution will contain two undetermined functions of the $x^i$, which we will take to be $A_u^{(0)}$ and $A_u^{(4)}$. As is usual with Fefferman-Graham expansions involving integer powers of $\rho$, we will also need to introduce logarithmic terms in the asymptotic expansion of $A_i$ in order to be able to solve the asymptotic equations of motion.

 

Next, we impose the constraint that $\hat A^2 =0$. To the first non-trivial order, this yields

\be
4 (A_\rho^{(0)})^2 + 2 A^{(2)}_i \hat A^{(0) i} - \hat g_{ij}^{(2)} \hat A^{(0) i} \hat A^{(0) j} =0
\ee
This  constraint can be rewritten in terms of $A_i^{(0)}$ and $\hat g^{(2)}_{ij}$

\be
4 (A_\rho^{(0)})^2  + \hat g_{ij}^{(2)} \hat A^{(0) i} \hat A^{(0) j}  - 2 \p_k A_\rho^{(0)} \hat A^{(0) k}=0 \label{nonloc}
\ee
and determines $A_i^{(0)}$ up to an overall multiplicative constant. Also,  
it implies a highly non-local relation between $A_i^{(0)}$ and $\hat g^{(0)}_{ij}$. In principle, one would need to check that also at higher order the equations of motion are compatible with the constraint. The check is facilitated by the fact that in purely $AdS_3$ spacetimes, the asymptotic expansion \eqref{fg} of the metric terminates at second order \cite{sksolod}. In any case, given that all we are considering are perturbations diffeomorphic to  Poincar\'e $AdS_3$, we do not believe that solving the constraint to all orders will pose any problem.

\bigskip

To summarize, the Fefferman-Graham-like expansion that we propose for the metric modes in asymptotically Schr\"{o}dinger spacetimes is the following:

\be
g_{\mu\nu} = \hat{g}_{\mu\nu} - A_\mu A_\nu \label{relads}
\ee
where $\hat{g}_{\mu\nu}$ is an AdS$_3$ metric with the usual Fefferman-Graham expansion \eqref{fg}, whereas the asymptotics of $A_\mu$ are given by \eqref{asya} where all the  constraints we derived are satisfied. This expansion coincides at linearized order with the expansion described in section \ref{linm}, provided we perform a diffeomorphism that brings $g_{\mu\nu}$ - rather than $\hat g_{\mu\nu}$ -to radial gauge.

\section{The holographic stress tensor \label{renact}}

We can now use our recently-developed asymptotic expansion to compute the renormalized action for the metric modes. Changing coordinates such that the $AdS_3$ metric takes the form
\be
ds^2 = d\eta^2 + \hat \g_{ij} \, dx^i dx^j
\ee
the metric of null warped $AdS_3$ can then be written as

\be
ds^2 = N^2 d\eta^2 + \g_{ij} (dx^i + N^i d\eta) (dx^j + N^j d\eta)
\ee
where

\be
N_i = - A_i A_\eta \;, \;\;\;\;\; N^2 =\frac{1}{1+A_\eta^2}\;, \;\;\;\;\; \g_{ij} = \hat \g_{ij} - A_i A_j
\ee
It follows that 

\be
\det \g = (1+A_\eta^2) \det \hat \g\;, \;\;\;\;\; \g^{ij} = \hat \g^{ij} + \frac{\hat A^i \hat A^j}{1+ A_\eta^2}
\ee
where we have manipulated the expressions using the fact that in $AdS_3$ the vector field is null

\be
\hat A^2 = \hat A^i A_i + A_\eta^2 =0
\ee
The extrinsic curvature of the surface $\eta = const$ is defined as

\be
K_{ij} = - \frac{1}{2N \ell} (\p_\eta \g_{ij}- D_i N_j - D_j N_i)
\ee
where $D_i$ is the covariant derivative compatible with $\g_{ij}$. In terms of $\hat\g_{ij}$ and $A_i$, $K_{ij}$ reads

\be
2 N \ell K_{ij} = -\p_\eta \hat \g_{ij} - \frac{A_\eta}{1+A_\eta^2} (\hat D_i A_j + \hat D_j A_i) + \frac{2 A^2_\eta}{1+A_\eta^2} (A_i \hat \e_{j}{}^k + A_j \hat \e_{i}{}^k) A_k
\ee
Using $\hat \nabla_\mu \hat A^\mu =0$, its trace can be simplified to

\be
2 N \ell K =  \frac{2\ell \hat K}{1+ A_\eta^2} - \frac{2 \hat A^i \p_i A_\eta}{(1+A_\eta^2)^2}
\ee
The on-shell action, which consists of a bulk term and a Gibbons-Hawking\footnote{We are assuming that the usual Gibbons-Hawking term renders the variational principle well-defined. Nevertheless,  this assumption needs further investigation.} term, reads

\be
S_{on-shell}=\frac{1}{16 \pi G} \int d^3 x \left( R + \frac{2}{\ell^2}- \frac{F^2}{4}- \frac{2 A^2}{\ell^2} \right) \sqrt{g} + \frac{1}{8 \pi G} \int K \sqrt{\g} 
\ee
Given that $F^2=A^2=0$, $R = -6 \, \ell^{-2}$ and $\det g = \det \hat g$, the value of the on-shell bulk action is the same as in $AdS_3$. Consequently, the regulated on-shell action can be written as

\be
S_{reg} =  S_{reg}^{AdS} - \frac{1}{8 \pi G \ell}  \int d^2 x \,\frac{\hat A^i \p_i A_\eta}{1+ A_\eta^2} \sqrt{\hat \g} \label{regact}
\ee
where 

\be
 S_{reg}^{AdS}  = \frac{1}{16 \pi G} \int d^3 x \left( \hat R + \frac{\,2}{\ell^2} \right) \sqrt{\hat g}  + \frac{1}{8 \pi G}  \int d^2 x  \,\hat K \sqrt{\hat \g}
\ee
The regulated $AdS_3$ on-shell action has an $\e^{-1}$ and $\ln \e$ divergence, whereas the extra term in \eqref{regact} has an additional $\e^{-1}$ divergence. It is well known that the $AdS_3$ divergence is cancelled by the counterterm 

\be
S_{ct}^{AdS} = - \frac{1}{8 \pi G \ell}\int d^2 x \sqrt{\hat \g} + \frac{\ell \ln \e}{32 \pi G} \int d^2 x \sqrt{\hat g^{(0)}} R[\hat g^{(0)}]
\ee
Consequently, the renormalized action  takes the form

\be
S_{ren} = S_{ren}^{AdS} +  \frac{1}{8 \pi G \ell }\int d^2 x \sqrt{\hat \g}  \left( 1- \frac{\hat A^i \p_i A_\eta}{1+ A_\eta^2}\right) + S_{ct} \label{sren}
\ee
The counterterm action is supposed to be a covariant expression in terms of the boundary fields $\g_{ij}$ and $A_i$, which cancels the above $\e^{-1}$ divergences. Noting that

\be
A_\eta = \sqrt{- \frac{ A^i A_i}{1+ A^i A_i}} \label{aeta}
\ee
we can write
\be
\sqrt{\hat{\g \,}} = \frac{\sqrt{\g}}{\sqrt{1+A_\eta^2}} = \sqrt{\g} \left( 1 - \frac{A_\eta^2}{2} + \frac{3 A_\eta^4}{8} - \ldots\right) =\sqrt{\g} \left( 1+ \frac{A^iA_i}{2} - \frac{(A^i A_i)^2}{8} + \ldots  \right) \label{ctgam}
\ee
which provides the first set of counterterms needed to cancel the first divergence completely. Importantly, the finite piece of the counterterms cancels exactly against the finite piece of the regulated on-shell action.  The same is true of the counterterms needed to cancel the second divergence, as $A_\eta$, according to \eqref{aeta},  
is a covariant function of the boundary data. Similar counterterms were previously encountered in \cite{ross}. 

\bigskip

In conclusion, the counterterm action $S_{ct}$  exactly cancels against the second term in \eqref{sren}, and the renormalized on-shell action for the metric modes is 

\be
S_{ren} = S_{ren}^{AdS}
\ee
where the $AdS_3$ spacetime is related to null warped $AdS_3$ via \eqref{relads}. It is thus very tempting to define the holographic stress tensor by

\be
T^{ij} = \frac{2}{\sqrt{- \hat g^{(0)}}} \frac{\d S_{ren} [\hat g^{(0)}]}{\d \hat g^{(0)}_{ij}} = \lim_{\e \r 0} \frac{2}{\sqrt{\hat \g}} \frac{\d S[\hat \g]}{\d \hat \g^{ij}}
\ee
with the result that

\be
T_{ij} = \frac{\ell}{8 \pi G} ( \hat g^{(2)}_{ij} - \hat g^{(0)}_{ij}\, \hat g^{(2)\,k}{}_k)
\ee
Consequently, just like in $AdS_3$, the stress tensor is conserved due to the asymptotic equations of motion and its trace \eqref{holoward} yields the conformal anomaly \eqref{cc}
which is identical to the one in pure Einstein gravity. Our result is consistent with those of \cite{kerrdip, dipCFT}, who were finding that the central extension of the asymptotic symmetry group of certain string-theoretical warped $AdS_3$ backgrounds equaled the one in the $AdS_3$ of the same radius. The asymptotic symmetry group can in principle be built as in $AdS_3$, by imposing Brown-Henneaux boundary conditions on the metric

\be 
\hat{g}_{\mu\nu} = g_{\mu\nu} + A_\mu A_\nu \label{bndm}
\ee
and appropriate boundary conditions on the massive vector field $A_\mu$, subject to the constraint $A^2=0$. We have not worked out the explicit boundary conditions and asymptotic charges, but we note that at least formally the Brown-York charges

\be
Q_\xi = \int \sqrt{\hat\s}\,  T_{ij} \,\hat\xi^i \hat n^j \, d\phi \label{by}
\ee
where  $\hat \xi^i$ are conformal Killing vectors of the boundary $\hat \g$ - herein assumed to be flat - are conserved. This fact follows as usual from  the conservation and tracelessness of the boundary stress tensor and the fact that $\hat \xi^i$ are conformal Killing vectors. The integral is performed over a timelike slice of the boundary with metric $\hat{\g}_{ij} = \rho^{-1} \hat{\eta}_{ij}$, and  $\phi = u+v$ is assumed to be compact. It would be very interesting to check whether this proposal for the asymptotic conserved charges makes physical sense and agrees with the usual definitions of conserved charges in asymptotically warped $AdS_3$ spacetimes. If this is true, then it would immediately follow that the asymptotic symmetry group of warped $AdS_3$ consists of two centrally-extended Virasoro algebras with central charge given by \eqref{cc}.

\section{Discussion \label{disc}}

In this article, we have presented the general asymptotic expansion and a rough holographic analysis for a particular set of modes present in null warped $AdS_3$ solutions of the three-dimensional Einstein-Proca theory. While our analysis sheds light on the relationship between certain quantities computed in null warped $AdS_3$ - such as the symplectic form and the holographic stress tensor - and their $AdS_3$ counterparts, it leaves many interesting questions unanswered. 

One such question is how to understand black holes in warped $AdS_3$ from a holographic point of view. Unlike black holes in $AdS_3$ or warped black holes in topologically massive gravity \cite{stromgirls}, warped $AdS_3$  black holes in massive vector theories are not locally diffeomorphic to the vacuum spacetime, and thus they are not described by the ``metric modes'' alone. Thus, in order to grasp them, we need to understand the general structure of the asymptotic solutions of the field equations in warped $AdS_3$, and how to perform holographic renormalization in the presence of both types of modes, $X$ and $T$. Also, we need to understand the generalization of our prescription \eqref{bndm} for constructing the boundary metric when the $X$-modes are present at non-linear level, as they are in the known black hole solutions. Another fact worth understanding is whether our construction can be generalized to higher dimensions, where the equation satisfied by the ``metric modes'' cannot be simplified to \eqref{eqna}. 

Another important point concerns the locality of the divergences - and thus of the counterterms used in holographic renormalization - as a function of the chosen boundary data, $\hat g^{(0)}_{ij}$. The counterterms that we used depend on the boundary vector field $A_i \propto A_i^{(0)}$, which bear a non-local relationship \eqref{nonloc} to $\hat g^{(0)}_{ij}$. Sometimes, the non-locality is just an artifact of our insisting that the boundary counterterms be commonplace covariant expressions  of the boundary data $\g_{ij}, A_i$. For example, while each term of the form $\sqrt{\g} (A^i A_i)^n$ in the expansion of the counterterm \eqref{ctgam} is non-local in terms of $\hat g^{(0)}_{ij}$, their sum is perfectly local. Otherwise put, if we require our counterterms to be commonplace-looking covariant expressions of the boundary field

\be
\hat \g_{ij} = \g_{ij} + A_i A_i
\ee
rather than $\g_{ij}$, then at least the first counterterm in \eqref{sren} would look entirely harmless. The equation above is simply a canonical transformation of the boundary data - provided the canonically conjugate momenta are transformed accordingly -
and it would be very interesting to check whether in the general case when $X$-modes are also included, holographic renormalization can be interpreted as a canonical transformation \cite{papadim} that asymptotically diagonalizes the map between the phase space and the space of solutions, while preserving the symplectic form. 

Returning to the issue of the locality of divergences, we would like to point out that the rules of holographic renormalization are not fully spelled out for spacetimes that 
are not asymptotically $AdS$, and it may be that in more general spacetimes the conditions that the boundary counterterms are required to satisfy are weaker. Indeed, it was found in \cite{baltmulti} that the boundary counterterms needed for the perturbative renormalization of irrelevant operators in AdS/CFT are generically non-local when expressed in terms of the boundary data, although they are local when expressed in terms of induced fields on the cutoff surface. Our counterterms are definitely local in terms of induced boundary fields at any given order in perturbation theory. 

Finally, it would be very interesting to connect our results to the asymptotic symmetry analyses mentioned in the introduction. For $AdS$ spacetimes it has been proven that the holographic Brown-York type conserved charges \eqref{by} equal those defined with the usual Lagrangian or Hamiltonian formalism up to a constant off-set which only depends on the boundary data \cite{comparison,compmarolf,skpap}. To understand the relationship between our formalism and the usual ones as far as computing the conserved charges is concerned, one not only needs to account for the boundary counterterms - a procedure which has been understood for various other spacetimes  \cite{compmarolf,compflat} - but also to integrate over a spatial slice of the boundary with metric $\hat \g_{ij}$ rather than $\g_{ij}$, which seems hard to motivate from a relativist's perspective. It is likely that the locality of the counterterms will be important in understanding whether the conserved charges agree in the two formalisms. In any case, if the $CFT_2$ structure we unraveled survives the inclusion of the $X$-modes and the two formalisms do turn out to be equivalent, this work may help us understand how to find boundary conditions which yield asymptotic symmetry groups consisting of two Virasoro algebras for warped $AdS_3$ spacetimes, and hopefully also for the near-horizon spacetime of the extremal Kerr black hole.

\bigskip

\noindent \textbf{Acknowledgements}

The author is very grateful to Geoffrey Comp\`ere, Kurt Hinterbichler,  Dan Jafferis, Denis Klevers, Gim Seng Ng and Balt van Rees for interesting discussions. The author would like to especially thank Andrew Strominger for encouragements and useful comments on the draft. 
 This work is supported in part by the DOE grant DE-FG02-95ER40893.

\end{document}